\documentclass[12pt]{iopart}

\usepackage{graphicx}
\usepackage{tikz,xcolor,hyperref}

\newcommand{\orcid}{\includegraphics[width=1em]{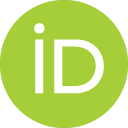}}

\begin{document}

\title[Understanding the working of a B-dot probe]{Understanding the working of a B-dot probe}

\author{Sayak Bose$^1$\footnote{Present address:
		Princeton Plasma Physics Laboratory, 100 Stellarator Road, Princeton, 
		 NJ 08540, USA.} \href{https://orcid.org/0000-0001-8093-9322}{\orcid{}}, Manjit Kaur$^1$\footnote{Present address:
		Department of Physics and Astronomy, University of California Irvine, Irvine, CA 92697, USA.} \href{https://orcid.org/0000-0001-6008-6676}{\orcid{}}, Kshitish K. Barada$^1$ \footnote{Present address:
		University of California Los Angeles, CA 90024 USA.} \href{https://orcid.org/0000-0001-7724-8491}{\orcid{}}, Joydeep Ghosh$^1$, Prabal K. Chattopadhyay$^1$ and Rabindranath Pal$^2$}

\address{$^1$ Institute for Plasma Research, HBNI, Bhat, Gandhinagar- 382428, India}
\address{$^2$  Saha Institute of Nuclear Physics, 1/AF Bidhannagar, Kolkata 700064, India}


\begin{abstract}
Magnetic pickup loops or B-dot probes are one of the oldest known sensors of time varying magnetic fields. The operating principle is based on Faraday's law of electromagnetic induction. However, obtaining accurate measurements of time-varying magnetic fields using these kinds of probes is a challenging task. A B-dot probe and its associated circuit are prone to electrical oscillations.  As a result, the measured signal may not faithfully represent the magnetic field sampled by the B-dot probe. In this paper, we have studied the transient response of a B-dot probe and its associated circuit to a time varying magnetic field. Methods of removing the oscillations pertaining to the detector structure are described. After removing the source of the oscillatory signal, we have shown that the time integrated induced emf measured by the digitizer is linearly proportional to the magnetic field sampled by the B-dot probe, thus verifying the faithfulness of the measured signal.
\end{abstract}


A plasma medium supports a large number of electromagnetic waves\cite{stix1992waves,swanson2003plasma}. Understanding the characteristics of these waves is essential for enriching our knowledge of plasma physics\cite{manjit} and for developing plasma based applications\cite{morita2017control}. One of the primary diagnostics used by plasma physicists to measure a time varying wave magnetic fields  is a B-dot probe\cite{Lovberg1965,hutchinson_2002}.    

A B-dot probe is typically a coil of conducting wire. The operating principle is based on Faraday's law of electromagnetic induction\cite{fewkes1965electricity}. An  electromotive force (emf), $V_{\rm ind}$, is induced in a coil when the magnetic flux threading through it changes in time. 
The magnitude of this induced emf is proportional to the time rate of change of the magnetic flux ($\Phi$), 
\begin{equation}\label{FL}
V_{\rm ind} \propto \frac{d \Phi}{dt} = N\frac{d \int \mathbf{B}. d\mathbf{A}} {dt}, 
\end{equation}
where, $\Phi=N\int \mathbf{B}.d\mathbf{A}$, $N$ is the number of turns in the coil, $\mathbf{B}$ is the magnetic field being measured and the integral is over the cross-sectional area of the coil. If the area of a B-dot probe is constant in time, then the average magnetic field sampled by the probe is given by, $B_{\rm avg}=\frac{\int \mathbf{B}.d\mathbf{A}}{\int d\mathbf{A}} =\frac{\int \mathbf{B}.d\mathbf{A}}{A}$. Therefore, Eq. \ref{FL} becomes
\begin{equation} \label{Eq:Bdot}
B_{\rm avg}=-\frac{1}{NA}\int V_{\rm ind}dt=C_{\rm calib}\int V_{\rm ind}dt. 
\end{equation}
Hence, as long as the calibration factor, $C_{\rm calib}$ is known, in principle, it is possible to obtain the value of the magnetic field from the induced emf. However, obtaining accurate measurements of magnetic field signals using B-dot probes are not as easy as its principle. The circuit  that is used to connect a B-dot probe to a digitizer for acquiring the data is prone to electrical oscillations that obscure the measurement. As a result, the integrated value of $V_{\rm ind}$ measured by the digitizer may not be linearly proportional to the magnetic field as predicted by Eq. \ref{Eq:Bdot}. Hence, it is a challenging task to interpret the probe signals correctly, when used to measure wave magnetic fields in complicated plasma devices such as Tokamak\cite{wesson1987tokamaks,ochoukov2015new,schittenhelm1997analysis}, Spheromak\cite{kaur_2018}, Stellarator\cite{wakatani1998stellarator}, etc.

In this paper, we present a simple inexpensive experiment to demonstrate the working of a B-dot probe and its associated circuit. Henceforth, we shall refer to a B-dot probe and its associated circuit as a B-dot probe circuit. The response of a B-dot probe circuit to a transient magnetic field is analyzed analytically and using circuit simulations. Two methods of removing sources of oscillatory signals from a B-dot probe circuit are presented. This article will be useful to students interested in learning the working of a B-dot probe as well as to experimental plasma physicists interested in measuring wave magnetic fields with high time resolution using a B-dot probe.

This article is organized as follows. The experimental set up is described in Sec. \ref{EXPS}. Oscillatory signals associated with transient magnetic field measurements using a B-dot probe are shown in Sec. \ref{Obs}. The identification of the source of the oscillatory signal is described in Sec. \ref{Source_identification}. Removal of the oscillatory signal  is demonstrated in Sec.  \ref{Osc_removal} followed by a discussion and summary in Sec. \ref{Disc} and \ref{summary}, respectively.

\section{Experimental set up} \label{EXPS}

\begin{figure*}
\includegraphics[scale=0.37]{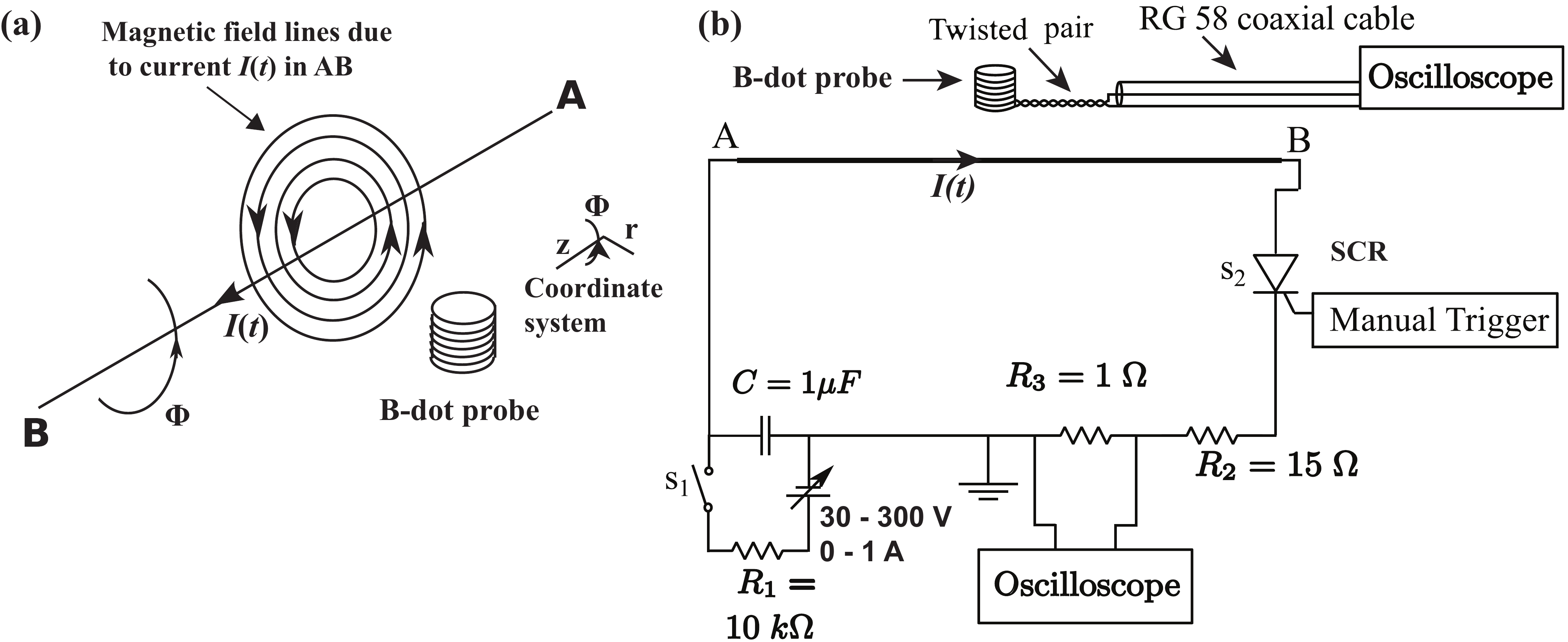}
\caption{\label{fig:fig_sch}(a) Long straight current carrying conductor AB with a B-dot probe placed beside it. (b) Schematic of the experimental set up.}
\end{figure*}

The experimental set up as shown in Fig.\ref{fig:fig_sch}a consists of a long straight current carrying conductor annotated as AB. A time varying current, $I(t)$ was passed through AB producing a time varying magnetic field, $\mathbf{B}(t)$ in its azimuthal plane. A B-dot probe placed beside AB samples $\mathbf{B}(t)$.

The circuit that was used to produce $I(t)$ is shown in Fig.\ref{fig:fig_sch}b. A  $\rm 1\mu F$ capacitor was charged using a power supply and a charging resistance, $\rm R_{1}=10 \;k \Omega$ by closing the switch $\rm s_{1}$. After charging the capacitor, the switch $\rm s_{1}$ was opened. The charged capacitor was discharged through AB, and resistances $\rm R_2 \left(=15\Omega \pm 20\%  \right)$ and $\rm R_{3} \left(=1\Omega \pm 20 \% \right)$ using another switch $\rm s_{2}$. The switch $\rm s_{2}$ was a semiconductor-controlled rectifier (SCR). The value of $I(t)$ following the closing of $\rm s_{2}$ was recorded by sampling the voltage drop across $\rm R_{3}$.

The B-dot probe used for measuring the time varying magnetic field was a single layered coil made using enameled copper wire of diameter $\rm 0.35 \pm 0.01\;mm$. The coil had $\rm 105$ turns, diameter, $2a=\rm 45.4\pm 0.1\; mm$, and a length, $b = \rm 39.4 \pm 0.1\; mm $. The end terminals of the coil are twisted tightly to form a twisted pair of length $\rm 11\;cm$. A $\rm 65\;cm$ long RG 58 coaxial cable was used to connect the twisted pair terminals of the B-dot probe to the oscilloscope as shown in Fig.\ref{fig:fig_sch}b.

\section{Initial Observations}\label{Obs}

\begin{figure}[t]
	\begin{center}
		\includegraphics[scale=0.6]{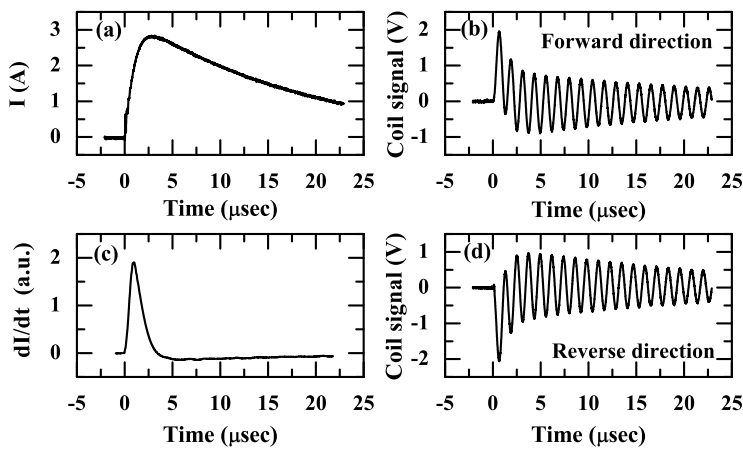}
	\end{center}
	\caption{\label{fig:FW_RD_I_dI} (a) Time variation of the current, $I$ in AB, (b) the B-dot probe signal as recorded by the oscilloscope, (c) first derivative of the current flowing through AB, and (d) the B-dot probe signal as recorded by the oscilloscope on reversing the direction of current through AB.}
\end{figure}

In the first step, the capacitor was charged to $\rm \sim 60\;V$. After charging, it was discharged through AB producing $I(t)$ as shown in Fig. \ref{fig:FW_RD_I_dI}(a). The signal acquired using the B-dot probe is shown in Fig. \ref{fig:FW_RD_I_dI}(b). Ideally, the B-dot probe output should be proportional to $dI/dt$. This is because, according to Biot Savart's law, $\mathbf{B}\propto I$. However, as shown in Fig \ref{fig:FW_RD_I_dI}(b) and (c), the B-dot probe signal is not proportional to $dI(t)/dt$.  An oscillation with a frequency of 820 kHz  appeared in the B-dot probe output, whereas there is no such oscillation present in $dI(t)/dt$. 

The questions that arise are (a) what is the source of this oscillation? and (b) Is there any coupling between AB and the B-dot probe ? A closer look at the B-dot probe signal reveals that the oscillations in the measured probe signal are initiated only when there is a temporal variation of the current in AB. Thus, it may be concluded that there is a definite coupling between AB and the B-dot probe. In order to establish the nature of the coupling between AB and the B-dot probe, the direction of the current in AB was reversed, thus reversing the direction of the generated magnetic field. The B-dot probe signal flipped as shown in Fig. \ref{fig:FW_RD_I_dI}(d) on reversing the direction of current in AB. This confirms that there is a definite magnetic coupling between AB and the B-dot probe. Hence, the B-dot probe is sensing the time varying magnetic field, but due to some other factors, it is exhibiting a damped oscillatory behavior. 

\section{Identification of the source of oscillations and its removal}\label{Source_identification}

\begin{figure*}[b]
	\includegraphics[scale=0.55]{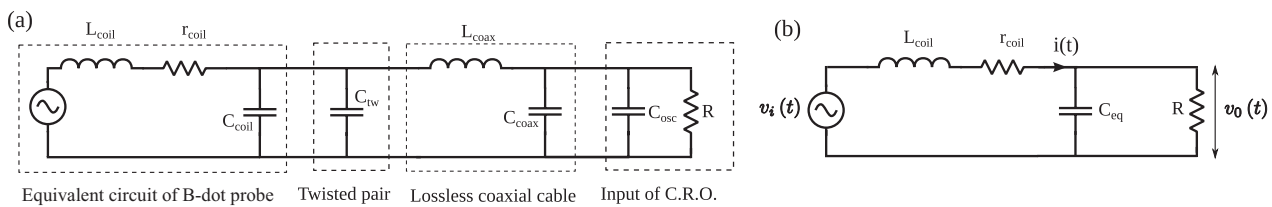}
	\caption{\label{fig:measurement_cir_osc}(a) Equivalent circuit representing the B-dot probe circuit. (b) Simplified equivalent B-dot probe circuit. }
\end{figure*}

Oscillatory signal of gradually decreasing amplitude normally arises in electrical circuits, consisting of an inductor, capacitor, and resistance (i.e. in LCR circuit) in under-damped conditions\cite{fewkes1965electricity}. In order to identify the source of inductance, capacitance, and resistance, each circuit component of the B-dot probe circuit is analyzed. 

The equivalent circuit of a B-dot probe, as shown in Fig. \ref{fig:measurement_cir_osc}(a), comprises of a voltage generator $v_{\rm i} \left(t\right)$ in series with an inductor and a resistor with a capacitor in parallel\cite{Segre_Allen}. The voltage generator, $v_{\rm i} \left(t\right)$ represents the emf induced in the B-dot probe by the changing magnetic flux. The inductor represents the inductance of the coil ($L_{\rm coil}$) and the capacitor represents the equivalent capacitance of the coil ($C_{\rm coil}$)\cite{feynman_V2}. According to Nagaoka's\cite{nagaoka,grover2004inductance} formula, the value of $L_{\rm coil}$ is $374\;\rm \mu H$. This calculated value is in good agreement with the measured value of $370\;\rm \mu H$. The measured resistance of the coil is $r_{\rm coil} = $ $3.2\; \rm \Omega$. The electrical parameters are measured using an LCR meter. The calculated value of $C_{\rm coil}$ using the formula given by Medhurst\cite{medhurst1947hf1} is $\sim 2\;\rm pF$.

The twisted pair connecting the B-dot probe to the coaxial cable (Fig. \ref{fig:fig_sch}b) is represented by a capacitor, $C_{\rm tw}$ as shown in Fig. \ref{fig:measurement_cir_osc}a. The measured value of $C_{\rm tw}$ is $\sim 12\; \rm pF$. The resistance is neglected as the length of the twisted pair is very small. The measured value of inductance, $L_{\rm tw}$ is $\sim 26 \rm \;nH$. $L_{\rm tw}$ is neglected as it is much small compared to $L_{\rm coil}$. In an LC oscillator, the energy oscillates between the electric field of the capacitor and the magnetic field of the inductor.  Since, $L_{\rm tw} \ll L_{\rm coil}$, the energy stored in the magnetic field of $L_{\rm tw}$ will be too small compared to the energy in the magnetic field of $L_{\rm coil}$. Therefore the effect of $L_{\rm tw}$ will be minimal in the observed oscillation. 

The 65 cm long RG58 coaxial cable is approximated as an ideal lossless coaxial transmission line. The capacitance of the coaxial cable ($C_{\rm coax}$) calculated by considering it to be a coaxial cylindrical capacitor is $\sim \rm 70.4 \;pF$. The inductance of the coaxial cable ($L_{\rm coax}$) determined using the formula for inductance of coaxial cylinders\cite{fewkes1965electricity} is $ \sim 0.17\; \rm \mu H$. 

The Tektronix made oscilloscope used for acquiring the data has a $\rm 20\; pF$ capacitor in parallel with $\rm 1\;M\Omega$ as its input termination as shown in Fig. \ref{fig:measurement_cir_osc}(a).

In order to analytically study the transient response of the measuring circuit, it is simplified further by considering only the dominant electrical components.  $C_{\rm coil}$ is neglected because it is very small compared to other capacitances in parallel to the B-dot probe. Since, $L_{\rm coax}\ll L_{\rm coil}$, the contribution of $L_{\rm coax}$ is minimal in the LC oscillations. Therefore, $L_{\rm coax}$ is also neglected. Thus, the simplified equivalent circuit of the B-dot probe (refer Fig. \ref{fig:measurement_cir_osc}b) consists of a voltage generator in series with $L_{\rm coil}$ and $r_{\rm coil}$ having a capacitor $C_{\rm eq}$ and resistance $R$ in parallel. $C_{\rm eq}$ represents the equivalent capacitance of the twisted pair, coaxial cable, and the input of the oscilloscope arranged in parallel $\left(C_{\rm eq} = C_{\rm tw} + C_{\rm coax} + C_{\rm osc}\rm = 12 + 70.4 + 20 = 102.4\; \rm pF\right) $.      

\subsection{Transient response of the simplified B-dot probe circuit}

\begin{figure}
	\begin{center}
		\includegraphics[scale=0.7]{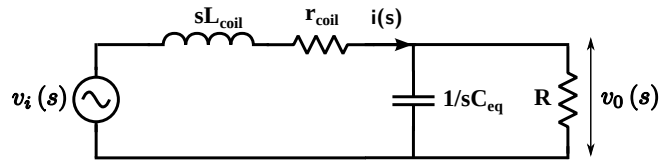}
	\end{center}
	\caption{\label{fig:LT}Laplace transform of the simplified B-dot probe circuit.}
\end{figure}

The transient response of the simplified equivalent circuit is studied by the Laplace transform technique\cite{spiegel1965schaum}. Fig. \ref{fig:LT} shows the transformed circuit. The impedance of the circuit as seen by the voltage generator is given by, 
\begin{equation}
z(s)=sL_{\rm coil} + r_{\rm coil} + \frac{R}{sC_{\rm eq}R+1}.
\end{equation} 

Therefore, the current through the inductor, $L_{\rm coil}$ is 
\begin{equation}
i(s)=\frac{v_{\rm i}\left(s\right)} {sL_{\rm coil} + r_{\rm coil} + \frac{R}{sC_{\rm eq}R+1}}.
\end{equation} 

The output voltage as observed in the oscilloscope, i.e. the voltage across the resistor is given by, 
\begin{equation}
v_{0}\left( s \right) = \frac{v_{\rm i} \left(s \right) \omega^{2}_{0}}{s^{2} + s \left(\frac{r_{\rm coil}}{L_{\rm coil}} + \frac{1}{CR} \right) + \omega^{2}_{0} \left( \frac{r_{\rm coil}}{R} +1 \right)},
\end{equation} 
where $\omega_{0}=1/ \sqrt{L_{\rm coil}C_{\rm eq}}$. A unit step voltage is applied in order to study the transient response of the circuit. In such a situation, $v_{i}(s)=1/s$ and the voltage drop across $R$ becomes 
\begin{equation}
v_{0}(s)=\frac{\omega^{2}_{0}}{s \left( s^2 + \alpha s + \omega^{2}_{0} \beta \right)}.
\label{eq:output}
\end{equation}

The output voltage in the time domain as shown in Eq. \ref{eq:conditions}, is determined by taking an inverse Laplace transform of Eq. \ref{eq:output} and performing some algebra.
\begin{equation}
v_{0} \left(t \right) = \frac{1}{\beta} - \frac{e^{- \psi t} \omega_{0}}{\sqrt{\beta} m } sin \left\{ mt + tan^{-1} \left(\frac{m}{\psi} \right) \right\},
\label{eq:conditions}
\end{equation} 
where $\beta = \frac{r_{\rm coil}}{R} + 1$, $m = \sqrt{ \omega^{2}_{0} \beta - \alpha^{2} / 4}$,  $\psi = \frac{\alpha}{2}$, and $\alpha = \frac{r_{\rm coil}}{L_{\rm coil}} + \frac{1}{C_{\rm eq}R}$. Critical damping, over damping and under damping corresponds to conditions $m^{2} = 0$, $m^2 < 0$, and $m^2 > 0$, respectively.

\begin{figure}[b]
	\begin{center}	
	\includegraphics[scale=0.5]{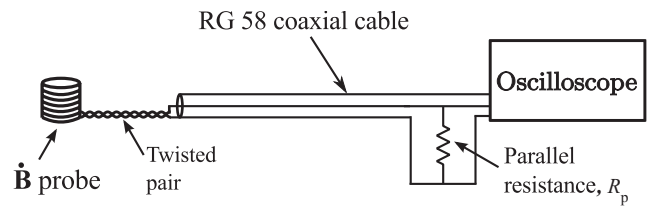}
	\end{center}
	\caption{\label{fig:Osc_rm_par} B-dot probe circuit with a parallel terminating resistance for removal of oscillatory signals.}
\end{figure}

The value of $m$ calculated using the experimental parameters is found to be $\rm 5.14\times10^6$ satisfying the under-damped condition $m^{2} > 0$. The calculated frequency of oscillations, $m/2\pi$ is 818 kHz, which is very close to the measured frequency of 820 kHz (refer Fig. \ref{fig:FW_RD_I_dI}). Therefore, it is concluded that the oscillations observed in Fig. \ref{fig:FW_RD_I_dI}(a) and (b) are under-damped LCR oscillations. Thus, in order to prevent the oscillations, the circuit needs to be over damped such that $m^2 < 0$.  The value of $m^{2}$ may be reduced by connecting a low resistance $R_{\rm p}$ in parallel to $R$ in the equivalent circuit shown in Fig. \ref{fig:measurement_cir_osc}(b). This is done by attaching the low  resistance $R_{\rm p}$ in parallel to the input of the oscilloscope as shown in Fig. \ref{fig:Osc_rm_par}. If $R_{p} \ll R$, then the effective terminating resistance is $R_{\rm p}$ and R is to be replaced by $R_{\rm p}$ in Eq. (\ref{eq:conditions}).

\begin{figure}[t]
\begin{center}	
	\includegraphics[scale=0.8]{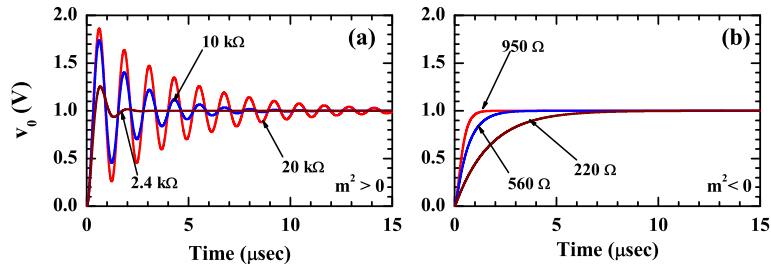}
\end{center}
	\caption{\label{fig:plot}Transient response of the B-dot probe circuit for different values of $m^{2}$ as given by Eq. \ref{eq:conditions}. Value of parallel terminating resistance is changed to vary $m^2$. }
\end{figure}

The analytical expression given by Eq. (\ref{eq:conditions}) is plotted for a set of $R_{\rm p}$ values for which $m^{2}$ changes from greater than zero to less than zero as shown in Fig. \ref{fig:plot}. We observe that as the value of $R_{\rm p}$ decreases, the oscillations die out, however, the rise time of the signal in the measuring circuit decreases for lower values of $R_{\rm p}$. The time resolution of the B-dot probe circuit is approximately represented by the time constant of an RL circuit\cite{fewkes1965electricity} given by $L/R$, which in this case is $L_{coil}/R_{\rm p}$. The time constant is the time in which the current in the circuit rises to $63 \%$ of its maximum value. If the characteristic rise time of the signal is less than the time constant, then the probe will not respond faithfully. 

\subsection {Effect of finite rise time of a signal in the B-dot probe circuit on the observed oscillatory behavior}

\begin{figure}[b]
	\begin{center}
		\includegraphics[scale=0.65]{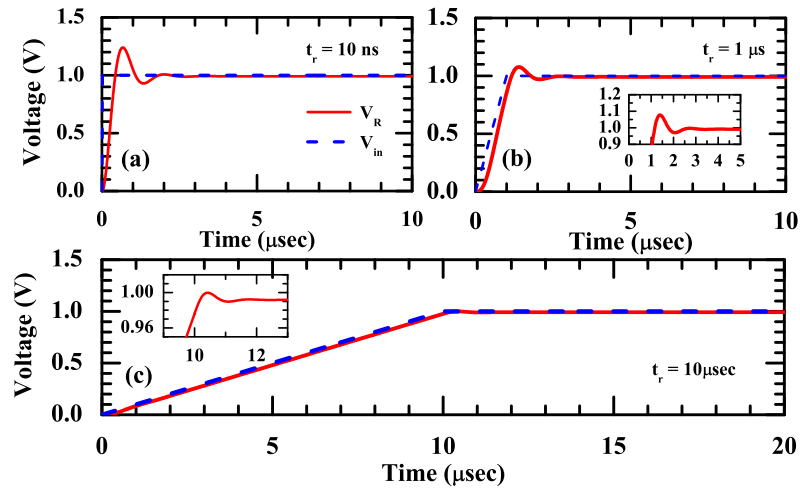}
	\end{center}
	\caption{\label{fig:simu}Effect of finite voltage rise time on the temporal response of the B-dot circuit.}
\end{figure}

The analytical calculations were carried out by applying a unit pulse voltage of infinitesimally small rise time to the equivalent circuit shown in Fig. \ref{fig:measurement_cir_osc}(b). In order to study the change in the oscillatory behavior of the measuring circuit with change in rise time of the voltage supplied by the voltage generator, the circuit in Fig. \ref{fig:measurement_cir_osc}(b) was simulated by applying an input voltage of the form
\begin{eqnarray}
v_{\rm i} \left(t \right) = \frac{t}{t_{\rm r}} \;\;\qquad\qquad  t\leq t_{\rm r}, \\
v_{\rm i}\left(t \right)=1          \;\;\;\;\qquad\qquad t>t_{\rm r},
\end{eqnarray} 
where $t_{\rm r}$ represents the rise time. 

A resistance, $R_{\rm p} = \rm 2.4\; k\Omega$, was used in parallel to $\rm R_{\rm osc}$ and the time evolution of voltage across it was simulated. The voltage across the $\rm 2.4\; k\Omega$ resistance was expected to exhibit  an oscillatory behavior as per the analytical result plotted in Fig. \ref{fig:plot}. In the simulations, the value of $t_{\rm r}$ was varied from $\rm 10\;ns$ to $\rm 10\; \mu s$. For the $\rm 10\; ns$ rise time case, the simulation results were in good agreement with the analytical results as shown in Fig. \ref{fig:simu}(a). But as the rise time ($t_{\rm r}$) of the applied voltage was increased, the amplitude of the oscillations decreased drastically as shown in Fig. \ref{fig:simu}(b)  and (c). Hence, for instantaneous rise, i.e. $t_{\rm r}\rightarrow 0$, the condition $m^2 \leq 0$ gives the resistance for removal of oscillations. For a finite rise time, a value of resistance for which $m^2$ is greater than zero may also minimize the oscillations.

\section {Demonstration of the removal of the observed oscillation}\label{Osc_removal}

\subsection{B-dot probe with a parallel terminating resistance}\label{parallel_resistance}

\begin{figure}
	\begin{center}
	\includegraphics[scale=0.65]{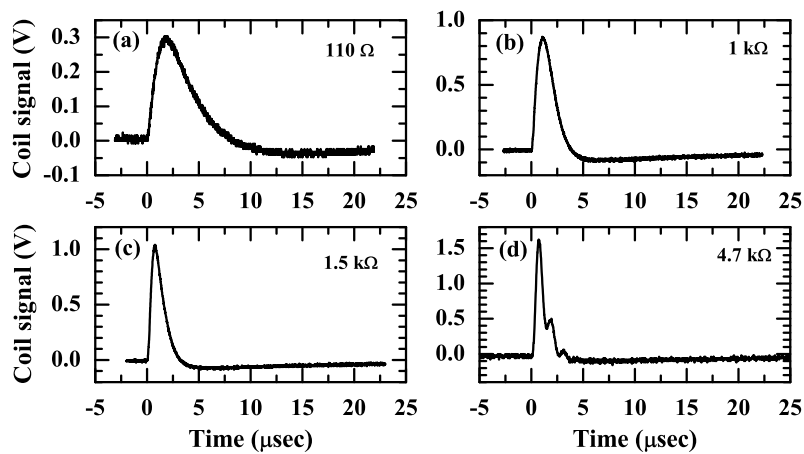}
	\end{center}
	\caption{\label{fig:par_res} B-dot probe output signal as measured in the oscilloscope for different values of parallel terminating resistances.}
\end{figure}

The effectiveness of a parallel resistance, $R_{\rm p}$ in removing the oscillations in the B-dot probe circuit was studied by varying the values of $R_{\rm p}$. 
The values of $R_{\rm p}$ were chosen to ensure that $m^2$ varies from less than zero to greater than zero. Fig. \ref{fig:par_res}(a), (b), and (c) demonstrate the removal of oscillations for three values of $R_{\rm p}$, 
$\rm 110\:\Omega$, $\rm 1\:k\Omega$, and $\rm 1.5\:k\Omega$, respectively. Values of $m^2$ are less than zero for both $\rm 110\;\Omega$ and $\rm 1\:k\Omega$, while for $\rm 1.5\:k\Omega$, it is greater than zero. Oscillations of the probe output signal were observed again, as shown in Fig. \ref{fig:par_res}(d), when $R_{\rm p}=\rm 4.7\:k\Omega$ $\left( m^2_{4.7k} > 0 \right)$ was used. The value of $m^2_{\rm 4.7k}$ is greater than $m^2_{\rm 1.5k}$ by 1.6 times. The non-observation of oscillations in the probe output for $R_{\rm p} = \rm 1.5\:k\Omega$ for which $m^2_{\rm 1.5k}>0$, may be due to finite rise time of $v_{\rm in} \left( \sim dB/dt \right)$ in the B-dot probe circuit.

\subsection{Alternate solution for removing the observed oscillations}\label{series_resistance}

\begin{figure}[b]
	\vspace{5 mm}
	\begin{center}
		\includegraphics[scale=0.55]{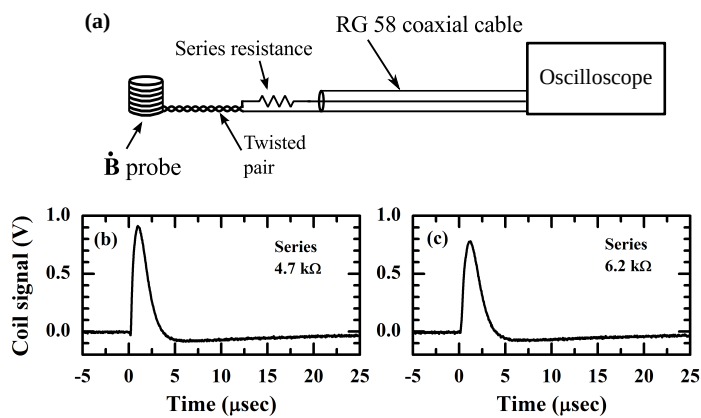}
	\end{center}
	\caption{\label{fig:series}(a) Circuit for measuring pulsed magnetic field using a resistance in series between the B-dot probe and the coaxial cable for removal of oscillatory signal. (b) B-dot probe output signal as measured in the oscilloscope for different values of series resistances. }
\end{figure}

In the B-dot probe circuit, the inductance is mainly contributed by the B-dot probe coil and the  capacitance is primarily contributed by the coaxial cable. Hence, it should be possible to remove oscillations in the measurement circuit by attaching a large resistance in series to the inductor as shown in Fig. \ref{fig:series}a, such that the measuring circuit effectively behaves as a critically damped or an over damped series LCR circuit. The series resistance is estimated by using the condition for over damping of LCR circuit $R_{\rm series}\geq \sqrt{4L_{\rm coil}/\left(C_{\rm cable}+C_{\rm osc} \right)}$. It was observed that indeed oscillations in the measuring circuit disappeared on attaching a large value of resistance in series to the B-dot probe coil as shown in Fig. \ref{fig:series}(b) and (c).

\section{Discussion}\label{Disc}

\subsection{Comparison of the two solutions for removal of oscillations }
B-dot probe circuit can be operated in either of the two configurations described in Sec \ref{parallel_resistance} and \ref{series_resistance} to remove LCR oscillations however, their susceptibilities to extraneous noise sources is different. A B-dot probe circuit with low parallel terminating resistance is less susceptible to noise than a circuit with a high series resistance\cite{ott1988noise}. Therefore, to ensure a high signal to noise ratio over a wide range of conditions, a low parallel terminating resistance is the preferred solution.   

\subsection{Choice of parallel resistance, $R_{\rm p}$}\label{guidelines}
 An estimate of the upper limit of $R_{\rm p}$ is given by the condition of no oscillation in a B-dot probe circuit, $m^{2} \leq 0$. While the lower limit of $R_{\rm p}$ is given by $f \ll 1/\tau$, where $f$ is the frequency of the signal being measured and $\tau\approx L_{\rm coil}/R_{\rm p}$ is the time constant of the B-dot probe circuit in overdamped condition. This is because if $f > 1/\tau$, then the signal will change before the B-dot probe circuit is able to respond as a result, the response will not be faithful. For the fastest probe response, the critical damping condition  $m^{2}=0$ should simultaneously satisfy $f\ll 1/\tau$.          

\subsection{Verification of the B-dot probe signal obtained using $R_{\rm  p}$}\label{verification}

\begin{figure}[t]
	\vspace{5 mm}
	\begin{center}
		\includegraphics[scale=0.45]{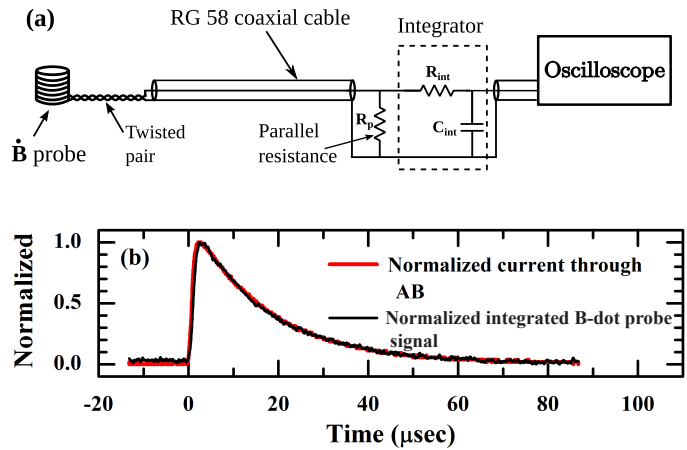}
	\end{center}
	\caption{\label{fig:int} (a) B-dot probe circuit with a passive integrator. (b) Superposition of the normalized time variation of current in AB and the integrated normalized B-dot probe signal.}
\end{figure}

After removal of the oscillations by following the guidelines in Sec. \ref{guidelines}, the fidelity of the probe signal was verified. The B-dot probe signal was integrated using a passive RC integrator as shown in Fig. \ref{fig:int}(a). The values of the resistance and capacitance of the passive integrator were $\rm 10 k\Omega$ and $\rm 80 nF$, respectively. The normalized integrated output of the probe signal was superposed on the temporal variation of normalized current in AB as shown in Fig. \ref{fig:int}(b). Both the signals were found to be in excellent agreement with each other. This shows that the measured magnetic field signal is linearly proportional to the current producing the magnetic field, thus confirming the faithfulness of the signal measured by the B-dot probe. The value of the magnetic field in units of Gauss can be determined by multiplying the integrated probe signal with a constant calibration factor, as the measured signal was varying linearly with the magnetic field. The method for determining the calibration factor is well described by Lovberg\cite{Lovberg1965}.

\section{Summary}\label{summary}

In this article, we have presented a simple table top experiment that does not involve the complexities of large scale plasma devices to demonstrate the working of a B-dot probe. A transient magnetic field was produced by discharging a capacitor through a straight current carrying conductor AB. The time varying magnetic flux was sampled by a B-dot probe placed in the azimuthal plane of AB. The pulsed magnetic field measurement circuit using a B-dot probe was found to be susceptible to signals of oscillatory nature. The time response of the B-dot probe circuit was studied analytically and by circuit simulation. The inductance of the B-dot probe and the capacitance of the coaxial cable used to connect the probe to the oscilloscope formed a tank circuit which we  established as the principal cause of the observed oscillations. More than one solutions for removing the oscillations are demonstrated. This is followed by validation of the fidelity of the measured signal by utilizing the linear relationship between the magnetic field and current that produces it. The B-dot probe circuit satisfies basic requirements of a good detector, faithful response, high signal to noise ratio, linear calibration curve, etc. Therefore, this experiment also serves as a model tabletop experiment to demonstrate how to design a good detector by removing systematic effects related to the detector structure. 
      
\section*{Orcid iDs}
\begin{itemize}
    \item Sayak Bose \href{https://orcid.org/0000-0001-8093-9322}{\orcid{}} \textcolor{blue}{\href{https://orcid.org/0000-0001-8093-9322}{https://orcid.org/0000-0001-8093-9322}}

    \item Manjit Kaur \href{https://orcid.org/0000-0001-6008-6676}{\orcid{}} \textcolor{blue}{\href{https://orcid.org/0000-0001-6008-6676}{https://orcid.org/0000-0001-6008-6676}}  
    
    \item Kshitsh Kumar Barada \href{https://orcid.org/0000-0001-7724-8491}{\orcid{}} \textcolor{blue}{\href{https://orcid.org/0000-0001-7724-8491}{https://orcid.org/0000-0001-7724-8491}}  
    
\end{itemize}

\section*{References}
\bibliography{iopart-num}

\end{document}